\documentclass[sn-mathphys,Numbered]{sn-jnl}

\usepackage{graphicx}%
\usepackage{multirow}%
\usepackage{amsmath,amssymb,amsfonts}%
\usepackage{amsthm}%
\usepackage{mathrsfs}%
\usepackage[title]{appendix}%
\usepackage{xcolor}%
\usepackage{textcomp}%
\usepackage{manyfoot}%
\usepackage{booktabs}%
\usepackage{algorithm}%
\usepackage{algorithmicx}%
\usepackage{algpseudocode}%
\usepackage{listings}%
\usepackage{siunitx}%
\usepackage{gensymb}%

\usepackage[nottoc,numbib]{tocbibind}%

\begin{document}

\title[Article Title]{Skillful neural network predictions of Saharan dust} 


\author*[1,3]{\fnm{Trish E.} \sur{Nowak}}\email{pn284@exeter.ac.uk}

\author[2]{\fnm{Andy T.} \sur{Augousti}}\email{augousti@kingston.ac.uk}

\author[3]{\fnm{Benno I.} \sur{Simmons}}\email{bsimmons.research@gmail.com}
\equalcont{These authors should be considered joint last author.}

\author[1]{\fnm{Stefan} \sur{Siegert}}\email{s.siegert@exeter.ac.uk}
\equalcont{These authors should be considered joint last author.}

\affil*[1]{\orgdiv{Mathematics and Statistics}, \orgname{University of Exeter}, \orgaddress{\city{Exeter}, \postcode{EX4 QH}, \country{UK}}}

\affil[2]{\orgdiv{Department of Mechanical Engineering}, \orgname{Kingston University}, \orgaddress{\city{London}, \postcode{SW15 3DW},\country{UK}}}

\affil[3]{\orgdiv{Centre for Ecology and Conservation}, \orgname{University of Exeter}, \orgaddress{\city{Penryn, \postcode{TR10 9FE}}, \country{UK}}}

\abstract{Suspended in the atmosphere are millions of tonnes of mineral dust which interacts with weather and climate. Accurate representation of mineral dust in weather models is vital, yet remains challenging. Large scale weather models use high power supercomputers and take hours to complete the forecast. Such computational burden allows them to only include monthly climatological means of mineral dust as input states inhibiting their forecasting accuracy. Here, we introduce DustNet a simple, accurate and super fast forecasting model for 24-hours ahead predictions of aerosol optical depth (AOD). DustNet trains in less than 8 minutes and creates predictions in 2.1 seconds on a desktop computer. Created by DustNet predictions outperform the state-of-the-art physics-based model on coarse $1^\circ$ x $1^\circ$ resolution at $95\%$ of grid locations when compared to ground truth satellite data. Our results show DustNet's potential for fast, accurate AOD forecasting which could transform our understanding of dust's impacts on weather patterns.
\newline

\textbf{Copyrights:} For the purpose of open access, the author has applied a ‘Creative Commons Attribution (CC BY) licence to any Author Accepted Manuscript version arising from this submission.}


\maketitle


\section*{Introduction}\label{main}
\addcontentsline{toc}{section}{Introduction}

The Earth’s atmosphere is loaded with $\approx26$ million tonnes of mineral dust - an atmospheric aerosol that represents the vast majority of mass burden in the atmosphere \cite{Gliss2021,Kok2023}. Each year, major sources emit $\approx5,000$ million tonnes of dust globally \cite{Kok2021improved} and, although the majority of this material sinks at source, a substantial portion is transported over vast distances \cite{vanderDoes2018}. Once in the atmosphere, mineral dust interacts with the Earth systems and impacts weather, climate, human health and infrastructure, from fisheries to aviation \cite{Shao2011, Knippertz2014, Highwood2014, Nenes2014, Miller2014, Jickells2014,Morman2014,Kok2023}.

Despite its importance, representing atmospheric dust aerosols in weather and climate models is challenging \cite{Parajuli2022, Kok2023}. For example, physics-based Numerical Weather Prediction (NWP) and climate models struggle to fully represent the dust cycle with adequate emission, transport and generation \cite{Evan2014, Kok2021, Gliss2021, Zhao2022}. Instead, the Integrated Forecasting System (IFS) of the European Center for Medium-Range Weather Forecasting (ECMWF) creates predictions that use aerosol optical depth (AOD) based on monthly-mean climatological fields only \cite{Bozzo2017}. A limitation in computational resources is highlighted as one of the reasons for the lack of a dedicated aerosol scheme, since such a development would significantly increase the computational burden of the system \cite{Mulcahy2014}. The monthly mean AOD, developed by the Copernicus Atmosphere Monitoring Service (CAMS), provides a reasonable trade-off in global weather forecasting. However, a more accurate representation of the AOD would have significant benefits, such as large improvements in the representation of the summer monsoon circulation or precipitation patterns in the Sahel region \cite{Bozzo2020, Balkanski2021}. 

Recent developments in the field of AI present a significant opportunity to overcome the computational burden of a dedicated physics-based aerosol scheme. Models such as GraphCast, Pangu-Weather, and FourCastNet can now skillfully predict the main ERA5 variables and in many cases outperform the state-of-the-art NWP models \cite{Lam2023, Bi2023, Pathak2022}. To date, attempts to forecast atmospheric aerosols with neural network architectures have shown varying levels of success. ``Satisfying” results were reported \cite{Kang2019, Daoud2021} when applying a long-short-term memory (LSTM) architecture to local AOD forecasts. The application of a U-NET architecture revealed a skillful detection of classified `dust events` at 67${\%}$ precision rate \cite{Sarafian2023}. A lack of comparisons to the current physics-based forecasts, or inclusion of standardised skill metrics, makes direct comparison between AOD forecasting models nearly impossible.

Here, we present a unique application of 2D convolutional neural networks (CNN) to forecast atmospheric aerosol levels. We use our model (hereafter `DustNet') to produce 24-hour spatial forecasts of AOD over North Africa. Computationally cheap and extremely fast, DustNet runs on a modestly configured laptop, rather than a high-power computer (HPC) - a fraction of the computational power required by traditional NWP models. The model trains in less than 8 minutes and predicts in 2.1 seconds. We compare the predictions of DustNet, and the corresponding daily CAMS forecasts, against the satellite-derived data using standard evaluation metrics, such as the root mean squared error (RMSE) and an accuracy correlation coefficient, to facilitate easy comparison with future AI models. The advantage of a smaller processing power requirement and rapid speed of prediction, combined with the accuracy of the forecast, makes our model a valuable complement to traditional AOD forecasting systems.


\section*{Results}\label{Results}
\addcontentsline{toc}{section}{Results}


\subsection*{DustNet model architecture and performance verification}\label{model archi}
\begin{figure}[h]%
\centering
\includegraphics[width=0.95\textwidth, trim={0.0cm 7.4cm 0.7cm 7.5cm},clip]{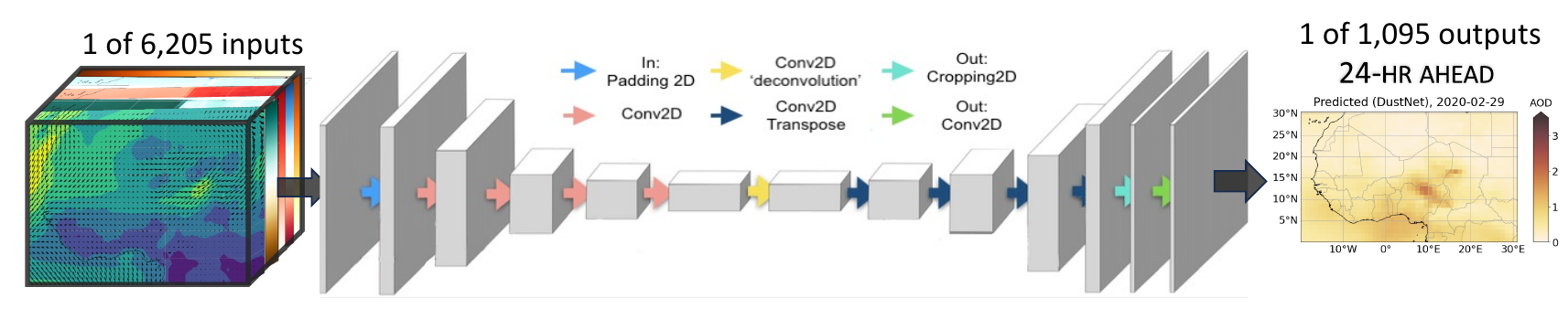}
\caption{Schematic representation of the DustNet model. Each of 6,205 inputs is first padded with a border of zeros using ZeroPadding2D (light blue arrow) to increase dimensionality and allow the convolution windows to detect the borders. The features are then extracted by 2D convolution window (pink arrows) which decreases dimensionality while increasing the number of trainable parameters. Then deconvolution is applied (yellow arrow) by including a 2D transpose network, which increases the size of the input (dark blue arrows) while maintaining connectivity between the layers. The output is then cropped back to match the initial input size (green arrow) and represents a 24-hr ahead prediction.}\label{fig1}
\end{figure}

\addcontentsline{toc}{subsection}{DustNet model architecture and performance verification}
To find the best deterministic AOD forecasting model, we compared three models. First, we adapted two leading CNN architectures, including 2-dimensional CNN and U-NET \cite{Hinton1995, LeCun2015, Goroshin2015unsupervised, ayzel2020}. We also custom-designed a 2D CNN with transposed layers \cite{Zeiler2010}. After comparing the performance of these three models (see \ref{table2:cnn-results} in the Methods section), we arrived at an optimal configuration for 24-hour dust aerosol forecasts and called our model DustNet. To train our DustNet model we used 17 years of daily AOD data (2003-2019) from the Moderate Resolution Imaging Spectroradiometer (MODIS) apparatus on board the Aqua and Terra satellites (see \nameref{methods} section for full details). A schematic representation in Fig. \ref{fig1} illustrates the inputs and output of the model. The inputs included the value of the AOD over the previous 5 days and previous 1 day for each of 35 meteorological features (7 atmospheric variables at 5 pressure levels, see \nameref{ERA5} section in \nameref{methods}). Regridded to a $1^\circ$x$1^\circ$ resolution over $31^\circ$ of latitude by $51^\circ$ of longitude, together with orography and the sine and cosine values of timestamps, the data resulted in a representative state consisting of 67,983 values for each training day. The compiled model yielded nearly 1.3 million trainable parameters and took 7min and 41s to complete the training process. Subsequently, the forecasts were produced in 2.1 seconds.

To evaluate the resulting 24-hour predictions, we used 3 years of data (2020-2022), which were unseen by the model. Our initial baseline model included the climatological mean, which is often used in meteorological forecasts as a sensible default \cite{Bozzo2020}. The baseline tests revealed that DustNet improved (reduced) the mean squared error (MSE) by $53.68\%$ in comparison to predictions based on the climatological mean. The regimes used for training, validation and testing are included in section \nameref{tvt}. To validate our results, we compared our predictions with the ground-truth (not imputed) data from MODIS, where the mean values between Aqua and Terra satellites, which record non-simultaneous measurements, provided the best representation of conditions around midday. To quantitatively assess the performance of the DustNet model against the ground truth we used two skill metrics: the root mean squared error (RMSE) and the anomaly correlation coefficient (ACC). To allow for comparison with the physics-based forecast, we tested the 24-hour predictions from CAMS using these same skill metrics, and compared them with the results produced by DustNet. 


\subsection*{Performance of spatial forecast}\label{rmse}
\begin{figure}[!ht]%
\centering
\includegraphics[width=0.97\textwidth, trim={0.5cm 3.7cm 0.5cm 3.4cm},clip]{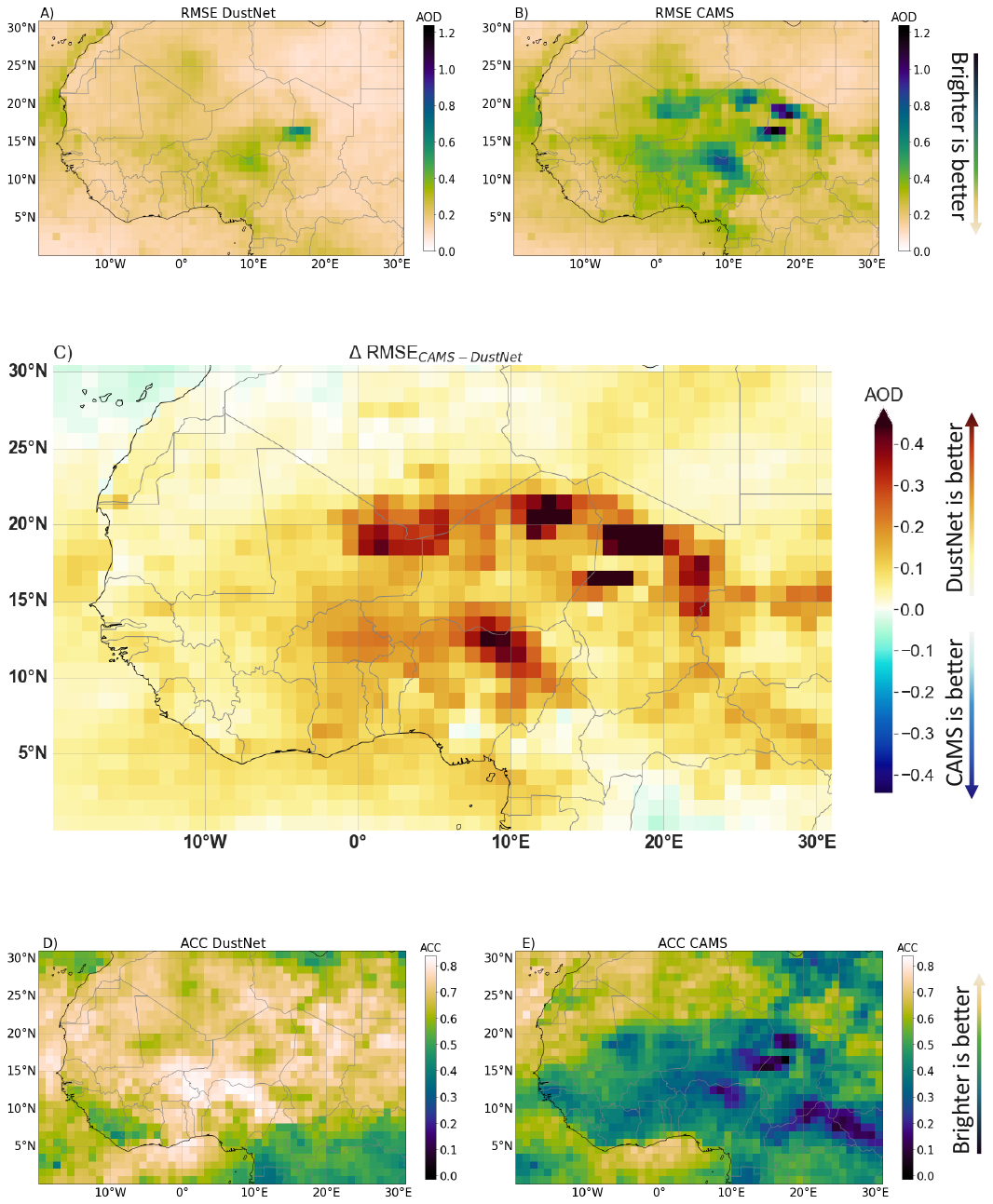}
\caption{Metrics indicating model performance. Results for 24-hour predictions of AOD values (2020-2022) compared with the ground truth data from MODIS. The RMSE for DustNet \textbf{(A)} and CAMS \textbf{(B)}, where the brighter the colour the smaller the error. Note, that the maximum error for DustNet is 0.62 AOD (medium green shades), while the maximum RMSE for CAMS reaches above 1.2 AOD (dark blue). In \textbf{C)} the difference in RMSE between CAMS and DustNet where all yellow to deep brown shades indicate the advantage of DustNet, while the blue shades indicate the advantage of CAMS. White grid cells indicate locations where both of the models performed equally when compared to the ground truth data. Note the lack of deeper blue shades and the dominance of yellow and brown grid cells where DustNet outperformed CAMS. \textbf{D)} and \textbf{E)} show the ACC for DustNet and CAMS respectively, where values above 0.6 (bright to white) indicate a valuable forecasting capability, while lower values (green to dark blue) indicate little to no predictive value. The ACC values in darkest blue indicate a misleading forecast. }\label{fig2}
\end{figure}

\addcontentsline{toc}{subsection}{Performance of spatial forecast}
We find that the DustNet model performs better in AOD forecasts than the physics-based CAMS model (Fig. \ref{fig2}). At nearly all spatial locations, DustNet predictions resulted in lower (better) RMSE values than CAMS during 2020-2022 (Fig. \ref{fig2}A and B). The greatest source of errors for both models was the most active dust source globally \cite{Todd2007} — the Bod\'el\'e Depression (16.5\degree N, 16.5\degree E). Although this is the location of the highest error, here we show again that DustNet’s RMSE is nearly $50\%$ lower than that produced by CAMS (0.62 \textit{versus} 1.24 respectively). The Bod\'el\'e Depression is of global importance for two main reasons: (i) it is responsible for over $50\%$ of the dust generated from the Sahara desert \cite{Todd2007, Washington2009, Jewell2021} and (ii) it was identified as the main source of minerals delivered seasonally to the Amazon basin \cite{Koren2006, Jewell2021}. A recent comparison of 14 physics-based models reveals their tendency to vastly underestimate the AOD forecast (ranging from -$16\%$ to -$37\%$) in comparison to ground-based observations \cite{Gliss2021}. With nearly 40 million tonnes of dust emitted annually from the Bod\'el\'e Depression, lowering the forecasting error at this location, as achieved by DustNet, has the potential to vastly improve the forecasting of transported dust.

Overall, DustNet predictions outperformed CAMS forecasts on $95.26\%$ of grid locations when comparing prediction errors (Fig. \ref{fig2}C). In Fig. \ref{fig2}C, grid cells in the darkest brown colour indicate locations where the errors produced by CAMS were over 0.45 AOD higher than that of DustNet, with the maximum error difference reaching 1.24 AOD. These locations represent central Saharan desert and arid regions, indicating the AOD composed of mineral dust, and thereby the more skillful ability of DustNet to capture dust generation. Moreover, DustNet captures the high mean AOD over northern Nigeria (associated with the seasonal Harmattan haze \cite{Anuforom2007, Sunnu2008, Schwanghart2008}) more skillfully than CAMS (details in section \nameref{seasonal} below). However, there are two locations at which CAMS forecasts performed better than DustNet (Fig. \ref{fig2}C). Both of these locations are adjacent to the boundaries (SE and NW corners), beyond which DustNet was unable to obtain information on the processes during training, while the data used to generate the CAMS forecast was extracted from a larger region (see Section \nameref{CAMS} for details). Thus, the lack of information on processes at the boundaries may have affected the CAMS forecasts less than it affected DustNet. This, however, might be overcome by extending the study region for DustNet.

We also compare the ability of DustNet and CAMS to detect anomalies using the ACC, a quantitative metric used in previous similar studies \cite{Lam2023, Bi2023} (see section \nameref{methods} for details). Here, DustNet also displays more skillful results than CAMS with a better (higher) ACC at $92.283\%$ of grid cells shown in Fig. \ref{fig2}D and E. An ACC score above $60\%$ is considered to be of value for forecasting purposes. The DustNet model surpasses this threshold at $79.89\%$ of locations (white-yellow), indicating a better forecast value for a wider range of locations than CAMS (which had an ACC value above $60\%$ at only $29.10\%$ of the grid cells). Skillful detection of anomalies, combined with a high forecast value, indicates that the DustNet model could be a valuable addition to Earth System Models, where better representation of Saharan dust events leads to more realistic forecasts of precipitation and a better representation of the African monsoon \cite{Anuforom2007, ECMWF2021, Balkanski2021}. 


\subsection*{Performance of seasonal-mean forecast}\label{seasonal}
\begin{figure}[!ht]%
\centering
\includegraphics[width=0.81\textwidth, trim={0.45cm 3.85cm 0.45cm 2.4cm},clip]{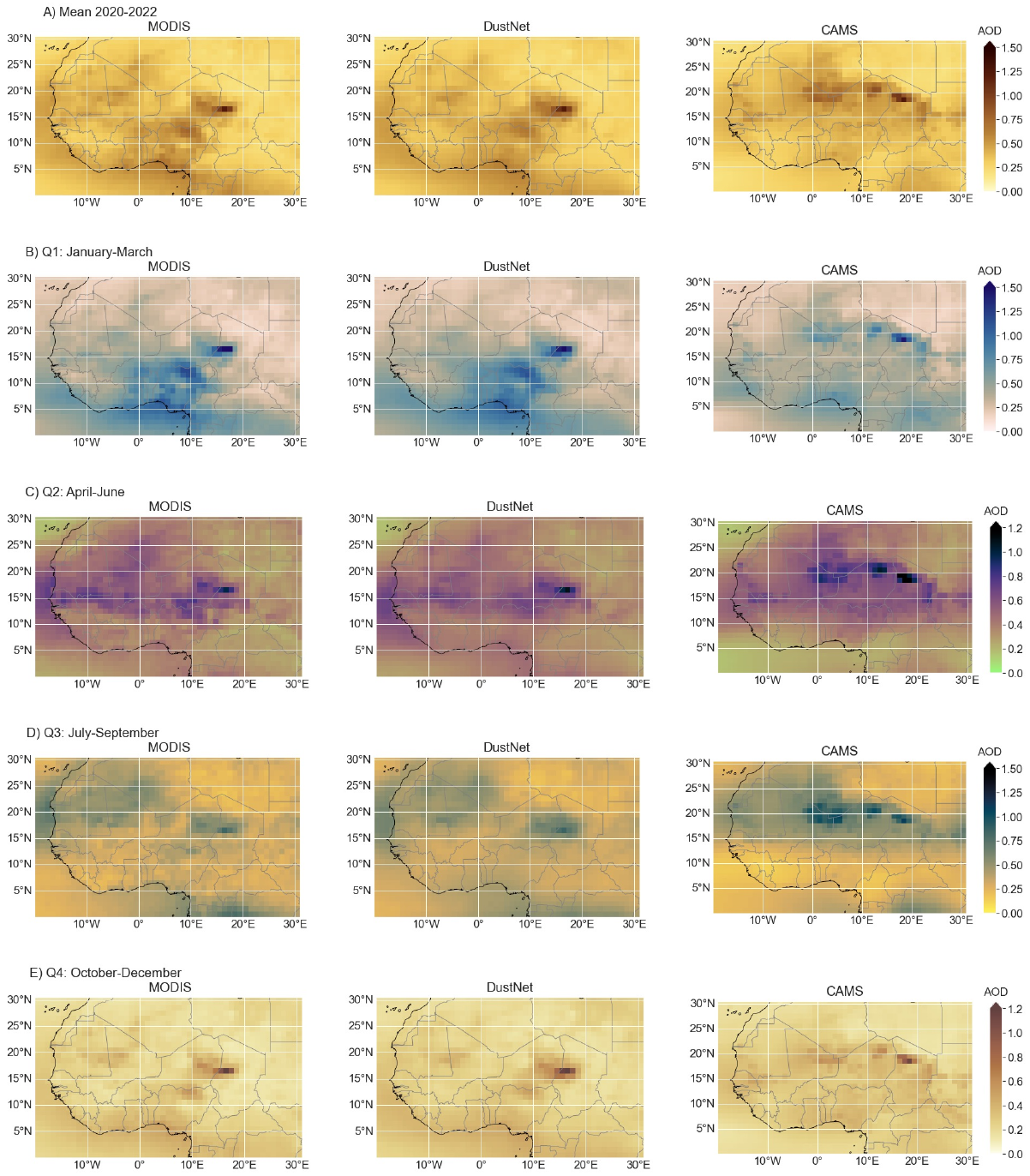}
\caption{Annual and quarterly mean AOD for 2020-2022. Mean AOD values calculated from 24-hr predictions. The left column represents AOD values from MODIS observations, while model predictions from DustNet are in the middle and from CAMS in the right column. \textbf{Row A)} compares the 3-year annual mean AOD between the observations and models, where DustNet skillfully captures the locations of the main dust events and the higher AOD around Nigeria and the Gulf of Guinea. In \textbf{row B)} the 3-year mean AOD for Q1: January - March, where the influence of the Harmattan wind has a visible effect on the mean AOD with a south-westward transport of mineral dust from the main generation site of the Bod\'el\'e Depression (dark blue). The effect of this transport is clearly picked up by our model. An increased AOD from biomass burning is also captured below 5\degree N. In \textbf{row C)} these same means are shown but for Q2: April - June where again the DustNet predictions skillfully capture the change in wind direction and westward aerosol transport in comparison to MODIS. \textbf{Row D)} shows that both models, CAMS and DustNet skillfully detected the northward shift of mean AOD transport during Q3: July-September. Here, CAMS forecasts tend to overestimate the AOD along the 20\degree\space latitude, but represent biomass burning related AOD around equator more realistically than DustNet, whose smoother contours seem to overestimate the AOD below 10\degree N. In \textbf{row E)} the seasonal decrease in aerosol activity for Q4: October - December is skillfully captured by both models when compared to observations from MODIS. Here, DustNet captures the position of the Bod\'el\'e Depression more accurately than CAMS and shows the lack of aerosol generation from the eastern locations. Note here the change in the colour-bar range.}\label{fig3}
\end{figure}

\addcontentsline{toc}{subsection}{Performance of seasonal-mean forecast}
Saharan dust aerosols are highly seasonal in emission and transport direction \cite{Anuforom2007, Schwanghart2008, Vandenbussche2020}. Therefore, here we additionally compared the annual and seasonal means of DustNet predictions with MODIS and CAMS. Fig. \ref{fig3}A shows the annual mean AOD values of MODIS and the model predictions. DustNet is capable of producing more realistic predictions in comparison to MODIS than the mean annual forecasts from CAMS. This is confirmed by a highly significant correlation of the annual spatial mean AOD (DustNet: r$^2$ = 0.91; CAMS: r$^2$ = 0.71, in Fig. \ref{figS1}). The DustNet model also captures the high AOD generated from the dustiest spot on Earth, the Bod\'el\'e Depression, more precisely than CAMS in both annual and all seasonal means (darkest colours in Fig. \ref{fig3}). Long-term comprehensive comparisons \cite{Gliss2021} show that the forecasts produced by physics-based models tend to underestimate the AOD values compared to ground observations. While this underestimation of AOD is clear between 5\degree N and 15\degree N, here we show that the CAMS forecast additionally tends to overestimate the AOD values around latitude 20\degree N over the Sahara during all the seasons of the period 2020-2022 (Fig. \ref{fig3}, rightmost panel and Fig. \ref{figS2}). This could be attributed to the locations of most of the ground observation stations, concentrated along latitude 10\degree N \cite{Gliss2021}. 

Moreover, we show that DustNet predictions capture the average seasonal displacement of AOD more skillfully than CAMS. The seasonal shift of Saharan dust by $\approx10$\degree\space in latitude is consistent with past observations and studies \cite{Prospero1981, Mbourou1997, Sunnu2008,Schepanski2017,Vandenbussche2020, Balkanski2021}. Comparisons of AOD in Fig. \ref{fig3}B and D indicate that DustNet captures this shift more skillfully than CAMS. Associated with a seasonal change in wind direction and large plumes of transported dust, this phenomenon is locally well known as the Harmattan haze and is responsible for the high increase in air pollution \cite{Anuforom2007, Schwanghart2008, Sunnu2008}. Previously noted mechanistic links between mineral dust and large-scale precipitation patterns, like the position of the Inter-tropical Convergence Zone (ITCZ) and the seasonal shift in the position of the West African monsoon, add to the importance of precise predictions of seasonal AOD displacement \cite{Sunnu2008, Janicot2008, NDatchoh2018, Balkanski2021}. Additionally, seasonal means of the AOD, extracted from short forecast lead times of reanalysis models including CAMS, are used to validate other models including climate models \cite{Zhao2022, OSullivan2020, Wu2020}. Thus, achieving higher accuracy for the predictions of seasonal mean AOD forecasts with DustNet could improve the performance of current forecasting models.

The smoothness of predictions displayed by DustNet in comparison to CAMS is a characteristic of the regression algorithm used by deep learning models (explained in \cite{Bi2023}).


\subsection*{Comparison of local predictions}\label{local}
\begin{figure}[!ht]%
\centering
\includegraphics[width=0.95\textwidth, trim={0.5cm 2.2cm 0.5cm 6.8cm},clip]{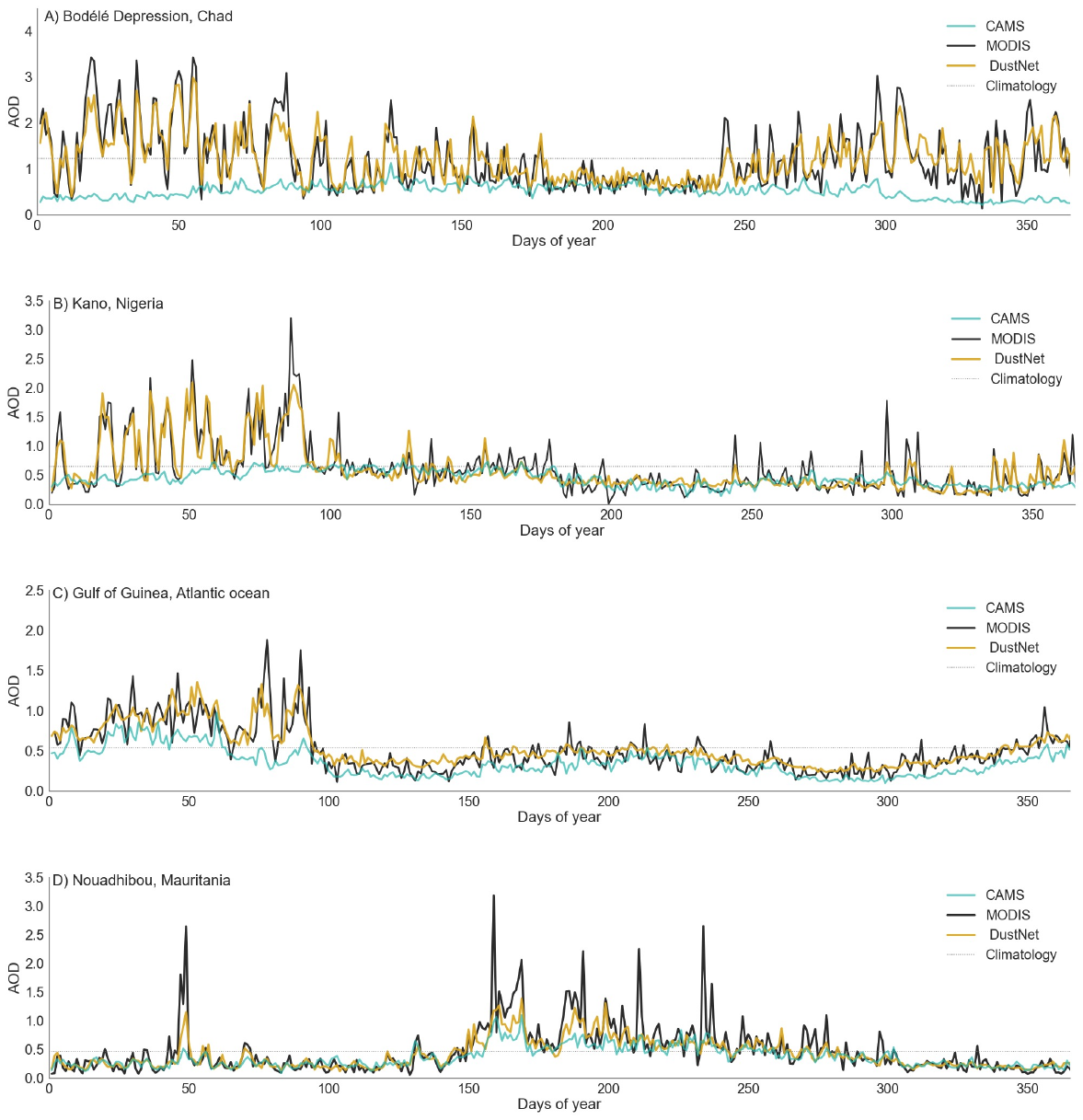}
\caption{Mean AOD predictions for each day of the year (2020-2022) at chosen locations. Shown are daily means (2020-2022) of AOD predictions from DustNet (golden line) and CAMS (light-sea-green line) as compared to MODIS (black line) and climatological mean (dotted line). At all four locations predictions from DustNet are closer to MODIS values than CAMS forecasts. An increase in AOD can be seen in the first ~90 days of the year in \textbf{A)} the Bod\'el\'e Depression, with lower but still elevated values towards \textbf{B)} Kano and \textbf{C)} Gulf of Guinea. These elevated AOD values during quarter 1 are not observed in \textbf{D)} Nouadhibou, which is consistent with the south-western direction of the Harmattan wind. DustNet also predicts daily and seasonal AOD variability at each site more skillfully than CAMS, whose forecasts tend to stay closer to or below the climatological mean. Both models struggle to fully capture the highest AOD peaks recorded by MODIS at the westmost location - Nouadhibou, however the DustNet model replicates these peaks better than CAMS. The background image, showing the position of the chosen locations (top), shows the December view of Blue Marble available from NASA https://visibleearth.nasa.gov/collection/1484/blue-marble?page=4.}\label{fig4}
\end{figure}

\addcontentsline{toc}{subsection}{Comparison of local predictions}
We also test the ability of DustNet to provide accurate 24-hr predictions at four locations indicative of the main dust transport routes (see methods for details on locations). At all four locations, DustNet predictions align with satellite data (MODIS) better than forecasts produced by CAMS (Fig. \ref{fig4}, and Fig. \ref{figS4} for correlations). This is especially evident at the Bod\'el\'e Depression, despite the site producing the highest prediction errors (see RMSE in Fig. \ref{fig2}A). The correlation between DustNet and MODIS is highly significant, with $r^2$ = 0.62, compared to CAMS which had $r^2$ = 0.01 (Fig. \ref{fig4}A and Fig. \ref{figS4}A). DustNet also skillfully detects the daily and seasonal variability of the Bod\'el\'e Depression, demonstrating the ability of our model to skillfully capture dust generation at this location. Similarly, 24-hr DustNet predictions for Kano, the second most populous city in Nigeria, align better with MODIS ($r^2$ = 0.74) than forecasts from CAMS ($r^2$ = 0.12), whose predicted values stay close to the climatological mean (Fig. \ref{fig4}B and Fig. \ref{figS4}B).

During the first quarter (DOY 0$\sim90$), the highest AOD values are present at the Bod\'el\'e Depression, Kano and the Gulf of Guinea (Fig. \ref{fig4}C). In Kano, the AOD values are just slightly lower than at the Bod\'el\'e and slightly lower in the Gulf of Guinea. Since both Kano and the Gulf of Guinea are positioned south-west from the Bod\'el\'e, their corresponding AOD values during quarter 1 indicate the Bod\'el\'e Depression as a generation source \cite{Schepanski2007, Jewell2021, Kok2021}. This also shows the ability of DustNet to capture generation and transport of AOD consistent with shifts in seasonal wind direction indicated in past studies \cite{Schepanski2017, Schwanghart2008, Anuforom2007, Sunnu2008}. During the third quarter (DOY 180$\sim270$), however, DustNet struggles to correctly capture the highest peaks in Kano and the Gulf of Guinea. The seasonal shift in meteorology and especially wind direction at these locations leads to an AOD composed of a mixture of aerosols, including sea-salt, and black carbon from biomass burning and industrial pollution \cite{Anuforom2007, Mari2008, Knippertz2017}. An area of future research could include information on vegetation and land cover during the training process, which would allow the model to distinguish between the ocean, Sahara Desert and central African forests. This would likely improve predictions for these regions and other aerosol species in general. The highest AOD values are also missed in Nouadhibou (Fig. \ref{fig4}D) during quarter 3 (DOY 180$\sim$260). However, here the seasonal increase in AOD points to a more localised origin, since dust generation at the Bod\'el\'e Depression is at its lowest with a daily AOD $\leq $1.0. This finding is consistent with past analyses of boreal summertime dust generation, which point towards Western Sahara, Mauritania, Algeria and Mali as dust sources \cite{Schepanski2007, Friese2017, Jewell2021, Kok2021}.


\section*{Discussion}\label{summary}
\addcontentsline{toc}{section}{Discussion}

The fast and skillful short-term predictions with DustNet present an opportunity for the forecasting community to incorporate a comprehensive aerosol scheme into future forecasts. The current coarse representation allows for quick testing and replication by professionals and enthusiasts alike. Despite DustNet not being explicitly trained to capture atmospheric processes such as dust generation, transport, or seasonal variations, these aspects are skillfully represented when compared to the satellite data. Furthermore, skillful representation of atmospheric aerosols at specific locations opens a possibility for DustNet integration into more localised weather models.

While DustNet outperforms CAMS in short-term forecasts, it is not without limitations. Although the model is trained on 43 features, only one - terrain - represented the ground conditions. Thus, incorporating additional information could be beneficial in capturing more nuanced or indeed wider interactions. For example, the generation of dust depends not only on the atmospheric conditions, but also on the soils mineral composition from which atmospheric dust derives \cite{Knippertz2017, vanderDoes2018}. Soil type and mineralogy impact the dust interactions with other atmospheric particles, and wider Earth systems by delivering essential minerals to the oceans and rain-forests \cite{Kok2023, Jickells2014, Koren2006}. Information on ground vegetation and cover can also play a role in determining dust generation locations and transport, especially over forests and in urban areas.

Additionally, DustNet’s predictions at the study region boundaries are visibly weaker than at the centre. This highlights the potential for a more skillful forecasts with a broader study area. Likewise, the predictions of extreme AOD values at point locations can fall short of the values captured by the satellites. Together with the deterministic nature of the model, DustNet’s predictions lack the probability distribution and the length of the tail for the extreme values.   

Addressing these limitations is crucial for future advancements. Rather than increasing the model's training time or epochs, we propose expanding the training data with diverse geographical information. This approach would capture nuanced interactions of atmospheric dust with Earth's systems. The inclusion of data from broader environmental disciplines, expanding study locations, and extending lead-time predictions are important next steps. Thus, a multidisciplinary approach can further enhance models capabilities and contribute to a range of specialised AI models with skillful predictions.

\section*{Methods}\label{methods}
\addcontentsline{toc}{section}{Methods}

\subsection*{Study area}\label{study area}
\addcontentsline{toc}{subsection}{Study area}
To effectively forecast dust aerosols, our study area encompasses the global principal dust generation source - the Sahara Desert - which is responsible for $\sim$55\% of the 1,536 million tonnes of total global dust emitted annually \cite{Ginoux2012}. The region (Fig. \ref{fig4} upper map) covers an area from 0\degree\space - 31\degree N and 20\degree W - 31\degree E ($31\times51$ grid cells), with a longitudinal centre around the Bod\'el\'e Depression (16.5\degree N, 16.5\degree E). Located in northern Chad, this single location generates an estimated 6--{18\%} of global dust emissions, which total to approximately \sisetup{separate-uncertainty}\num{182(65)} million tonnes per year, the region is of major importance in models that seek to capture dust generation \cite{Todd2007}. To capture the seasonal south-westward dust transport across the Sahara and towards the Atlantic Ocean, our region includes additional grid cells to the south and west of the Bod\'el\'e Depression.


\subsection*{Datasets}\label{Datasets}
\addcontentsline{toc}{subsection}{Datasets}

\paragraph*{AOD data}\label{AOD}
We retrieved the AOD data from the Moderate Resolution Imaging Spectroradiometer (MODIS) instrument located on board both Aqua and Terra spacecraft. With daily temporal resolution over a period of 20 years starting from 1st January 2003 to 31st December 2022, the AOD data yields $2\times7305$ files. We used the quality-controlled level-3 data for AOD at 550nm. Choosing the combined mean of Dark Target and Deep Blue algorithms provided a full coverage above bright and dark surfaces at a horizontal resolution of $1\degree \times 1\degree\space$ \cite{Hubanks2015}. This choice provided a good spatiotemporal coverage of AOD data above both land and ocean surfaces.

\paragraph*{ERA5 data}\label{ERA5}
Meteorological data comes from the fifth generation of European Centre for Medium-Range Weather Forecast (ECMWF) atmospheric reanalysis project (ERA5) and consists of 5 parameters: wind u component, wind v component, vertical velocity, temperature and relative humidity. Each parameter was retrieved at 5 pressure levels 550hPa, 750hPa, 850hPa, 950hPa and 1000hPa. This choice provided us with 35 distinctive features representing atmospheric conditions from ground level to $\approx5$km in vertical height. The ERA5 data is available on an hourly basis, but here we only chose the data representing conditions for midday (12:00 UTC). This allows us to represent the mid-point in atmospheric conditions between the Terra and Aqua satellite overpasses above the equator (10:30am and 1:30pm respectively). To further match the meteorological data with AOD, we chose a daily temporal resolution between 2003-2022. The horizontal resolution of ERA5 data is $0.25\degree \times 0.25\degree$. To match this with the AOD resolution of $1\degree \times 1\degree\space$, the data was regridded (see section \nameref{regrid} for details).

\paragraph*{Timestamps}\label{Timestamps}
 We created timestamps using the NumPy package (version 1.23.0) in Python with a daily temporal resolution over 20 years from 2003 to 2022 (7,305 days). Then we multiplied the file to match the exact spatial resolution of atmospheric variables and a coverage of $31 \times 51$ grid cells for each day.

\paragraph*{CAMS forecast}\label{CAMS}
We obtained daily `Total aerosol optical depth at 550nm' forecast data from `CAMS global atmospheric composition forecasts'. CAMS forms a part of the ECMWF Integrated Forecasting System (IFS), and is a sophisticated numerical weather forecasting model (NWP) \cite{Bozzo2017}. During the AOD data assimilation process, CAMS utilises data from MODIS, among other satellites, together with data from ground-based observation stations. The model then uses physics and chemistry principles to forecast hourly AOD values on a single level for up to 5 days (120hr) ahead \cite{Morcrette2009, Benedetti2009}. For consistency, we only chose forecasts representing 12:00 UTC to capture the midpoint conditions between Aqua and Terra overpasses above the equator. The temporal extent choice was also matched to our predictions. Therefore, we initiated forecasts on midday 1st January 2020 until 30th December 2022 for 1095 days forecast between 2nd January 2020 and 31st December 2022. CAMS data is provided at a $0.4\degree \times 0.4\degree\space$ spatial resolution. To match with our data, we therefore used an identical approach as for the ERA5 datasets to regrid to a $1\degree\times 1\degree\space$ resolution (details in \nameref{regrid} section). 


\subsection*{Data pre-processing}\label{regrid}
\addcontentsline{toc}{subsection}{Data pre-processing}

\paragraph*{Data imputation}
We combined data from the MODIS Aqua and Terra data sources at each individual location and time by labelling AOD data as missing whenever both sources were missing, using available data from one source if the other is missing, and averaging both sources whenever both are available. This data combination step reduces the total fraction of missing AOD values from $32.81\%$ in Aqua and $30.89\%$ in Terra to $19.89\%$ in the combined data set. The remaining missing AOD values are imputed by spatial interpolation (individually for each time step) using Lattice Kriging \cite{hartman2008, Rue2005} on four nearest neighbours with uniform weights. To validate the imputation method, we randomly held out $10\%$ of the AOD data and compared them to their imputed values. The mean squared error of the imputed values is $0.005$ which is less than $5.30\%$ of the total variance of the AOD data. The MSE was found to be insensitive to the choice of the Kriging hyperparameter, with relative differences of less than $0.0003\%$ over a wide range of values (see supplementary Fig.\ref{figS5}). See \nameref{links} section for links containing the full Python code for imputation.

\paragraph*{AOD lag} We use 5 preceding days of imputed AOD data as features to predict AOD on a given day. Hence, we had to remove the first 5 timestamps from the database as these did not have complete features available, resulting in a new total of 7,300. 

\paragraph*{ERA5 regridding} The ERA5 data \cite{Hersbach2018} is supplied with a vertical resolution of $0.25\degree \times 0.25\degree$ and thus needed regridding to match the AOD resolution. We processed all meteorological data using Python version 3.8.13 and the Iris v 3.2.1 package. We used nearest-neighbour interpolation from the Iris package to convert each feature to a common $1\degree\times1\degree$ resolution.

\paragraph*{Combining and normalising} We combined the meteorological data with AOD data into a single 4D NumPy array of shape {7300, 51, 31, 41}, where the first dimension represents time, the second and third are longitude and latitude respectively, and features are stored along the last dimension. Let the $x\textsubscript{ijt}$ be the value of feature $x$ at grid point $i,j$ and time $t$. We normalised all features using min-max normalisation:
\begin{equation}
    x_{ijt,norm}=\frac{(x_{ijt} - x_{min})}{(x_{max} - x_{min})}
    \label{eq:normalisation}
\end{equation}
where x\textsubscript{min} and x\textsubscript{max} are the overall minimum and maximum of a feature $x$ over all grid points and timestamps in the training data.

\paragraph*{Seasonal features}
Our first 41 features contain atmospheric variables as described above. Additionally we included the sine and cosine of timestamps as seasonal features using: 
\begin{equation}
    x^{(42)}_{ijt}=\sin\left(2\pi \frac{t}{365.2425}\right),
    \label{eq:tsin}
\end{equation}
and similarly using the cosine:
\begin{equation}
    x^{(43)}_{ijt}=\cos\left(2\pi \frac{t}{365.2425}\right)
    \label{eq:csin}
\end{equation}
Timestamps are constant across space and allow the model to represent periodic variations on seasonal timescales. Thus, together with timestamps, our final total input consisted of 43 features.
 

\paragraph*{Training, validation, test split}\label{tvt} We split the data along the time dimension into {70\%}, {15\%} and {15\%} for training, validation and test sets respectively. Splitting data with consecutive time steps yielded better results than a random split. Therefore, the training set covered 5,110 consecutive days from 6th January 2003 until 1st January 2017 (inclusive of both days). The validation set took 1,095 consecutive days from 2nd January 2017 to 1st January 2020. Finally, we set aside a test set, with 1,095 days of data from 2nd January 2020 to 31st December 2022. We made sure that the model never had access to the test set during the training and validation processes and only after these were complete did we introduce the test data and run our model to obtain predictions.


\subsection*{Designing CNN models}\label{CNN-models}
\addcontentsline{toc}{subsection}{Designing CNN models}

To find the best forecast of the daily AOD, we designed three CNN models based on \cite{Hinton1995, LeCun2015, Goroshin2015unsupervised}. We used the end-to-end open source machine learning platform TensorFlow 2, together with the Keras high-level API \cite{Chollet2015keras}. Each model uses a different architecture based on two-dimensional (2D) convolutions (hereafter Conv2D). In general, the Conv2D neural network architecture enables regression problems in image analysis to be addressed and is particularly effective at capturing spatial patterns in two-dimensional images. The efficiency of Tensorflow allows for training and inference to be run on traditional desktops or laptops rather than requiring HPC’s. All models described hereafter were run using Python version 3.10.10 on a MacBook Pro with an Apple M1 Pro and 32GB RAM.

We also performed a series of diagnostic tests in order to choose the best optimizer and loss function. Tested loss functions included CoSine, Huber, LogCosh, Mean Absolute Percentage Error, Mean Absolute Error, Mean Squared Logarithmic Error and Mean Squared Error. We assessed the performance based on the lowest mean squared error, the speed of the overall training time and the time taken per step. The mean-squared-error (MSE) loss function together with the Adam optimizer offered optimal results and was used for further analysis. For the Adam optimizer we used a learning rate of 0.001 and an exponential decay rate of 0.9, which are default settings following \cite{Kingma2014}. 

We determined the optimal size of the convolving window (kernel size) and the number of strides with a series of diagnostic tests. The results of these tests are presented in Table \ref{table1:kernel-size} with the optimal choice in bold based on minimising the mean squared error and the speed of the training time. The final design included a kernel size of (2,2) with a stride equal to 2, which produced the optimal MSE to training time ratio. We recognise that we have not tested every possible combination, thus a it may be possible to achieve a better performing design. 

\begin{table}[h]
\caption{Test results of choosing different kernel sizes for 2 models: Conv2D.T and U-NET. For simplicity, this test was run on a subset of data. The optimal choice is presented in bold font. Note that a small improvement in the MSE for a kernel size (3,3) was disregarded in favour of a much faster training time and time per step for kernel size (2,2).}\label{table1:kernel-size}
\begin{tabular*}{\textwidth}{@{\extracolsep\fill}lcccccc}
\toprule%
& \multicolumn{3}{@{}c@{}}{Conv2D} & \multicolumn{3}{@{}c@{}}{U-NET} \\\cmidrule{2-4}\cmidrule{5-7}%
Kernel size & (5,5) & (3,3) & (2,2) & (5,5) & (3,3) & (2,2) \\
\midrule
Training time & 42min & 23min & \textbf{3min} & 1h41min & 1h7min & \textbf{23min}\\
Time per step & 28s & 28s & \textbf{12s} & 144s & 140s & \textbf{88s}\\
MSE & 0.00174 & 0.00133 & \textbf{0.00134} & 0.00175 & 0.00148 & \textbf{0.00151}\\
\botrule
\end{tabular*}
\end{table}

We initially assigned 50 epochs to each training regime and monitored the performance using the mean squared error of training to validation loss. We also configured each model with Early Stopping and a patience of 4 epochs. This set up halts the training time when there is no improvement in validation loss after 4 consecutive iterations and prevents the model from over-fitting to training data (see supplementary Fig. \ref{figS8:train-valid}). Our set-up saved the optimal ratio of training time versus validation loss and used the best performance to run predictions. Below, each model’s architecture is described in detail.

\paragraph*{Conv2D model} For the first AOD prediction model we adapted a classical design of CNN. The Conv2D architecture, inspired by the visual system, applies filters (or convolutions) to capture spatial patterns in two-dimensional images \cite{LeCun2015}. The network performs feature extraction and learns representations at different scales. Such representations allow the network to identify relevant information and thus make predictions. Each of the hidden layers in our model was designed with a maximum of 264 and a minimum of 16 filters, as well as a $2\times2$ kernel size, which specifies the height and width of the 2D convolution window. Learning of the complex representation is made possible by the non-linearity provided to the model by a correctly chosen activation function. Ramachandran et al., \cite{Ramachandran2017} suggested an improvement to the popular ReLu activation function by proposing the Swish function. This method gained in popularity as it is capable of smoother output representation as well as more consistent performance \cite{Rasamoelina2020}. The Swish activation function proved to yield the best performance and thus we used it throughout the model layers. An architecture constructed in this way provided 1,291,009 trainable parameters.

\paragraph*{U-NET inspired model}The architecture of our second model employed a U-NET like design, first proposed by \cite{Ronneberger2015} for the purpose of biomedical image classification. The model is characterised by its “U” shape design which employs both contracting and expanding pathways to identify specific features within images. Here, we follow the approach of \cite{ayzel2020} who, inspired by U-NET, designed their RainNet model for precipitation nowcasting. Thus, we also divided our model into two parts, encoder and decoder, and utilised skip connections between both paths via concatenation layers - unique features of the U-NET model. The encoder (or contracting) pathway of the model included six Conv2D layers with Swish activation and a $2 \times 2$ kernel size, as well as two MaxPooling2D layers with pool size $1 \times 1$. The decoder (or expanding) pathway had five Conv2D layers with two UpSampling2D and two Concatenate layers. The input layers were bordered with a ZeroPadding2D layer which was cropped to the original size of $31\times 51$ with Cropping2D in the output layer. Unlike the original U-NET network, our design received 4-dimensional arrays of shape $7,300 \times 31 \times 51 \times 42$ and generated an output image of a shape of $31 \times 51$ for each prediction time step. Thus, the prediction generated 1905 images corresponding to dates from 2nd January 2020 to 31st December 2022.

\paragraph*{DustNet model} The last model design built upon the architecture of Conv2D and U-NET. This unique design replaces the Concatenation layers with Transpose convolution layers, also known as Deconvolutional Networks \cite{Zeiler2010}. Schematically represented in Fig. \ref{fig1} the input layer was first padded with a border of zeros (ZeroPadding2D) which increased the input shape from $31 \times 51$ to $40 \times 64$. Zero padding enabled the convolution to produce the same output size for multiple input sizes \cite{Dumoulin2016}. We then applied the 2D convolving windows (Fig. \ref{fig1} - pink arrows) which moved over each padded input with a $2 \times 2$ kernel size and $2 \times 2$ strides which allow upsampling. This allowed the model to decrease the input size (down to $5 \times 8$) while increasing the amount of trainable parameters. A ‘deconvolution’ was then applied by adding Conv2D Transpose layers. An advantage of transposed convolution is its ability to efficiently upscale input data by applying inverse convolutions. This enables the network to increase the size compared to the input and thus generates high-resolution images at finer spatial scales \cite{Zeiler2010}. A 2D cropping layer was then added to bring the shape back to its initial input size of $31 \times 51$. The final architecture allowed the model to create a total of 1,286,913 trainable parameters. Since this design yielded the optimal results of predicting dust aerosols in comparison to baseline models, we called it DustNet. 

\paragraph*{Baseline models} We set the baselines as AOD climatological mean and persistence. The climatological means were calculated separately at each spatial location as the mean AOD over the training period. The climatological benchmark is constant in time. A time-varying baseline model is the persistence forecast, which uses the most recent observation of AOD as the 24-hour ahead prediction. Here, we used the values from the 1st day of calculated AOD lag from the reserved test set (values unseen by the model) to represent persistence. Both climatology and persistence act as null models, and a more sophisticated forecasting scheme should be able to outperform both in order to be considered useful.

\subsection*{Statistical analysis}
\addcontentsline{toc}{subsection}{Statistical analysis}

\paragraph*{CNN models evaluation}\label{model-evaluation}
We evaluated each CNN model’s performance by assessing the training time, inference time taken per ‘time-step’, the MSE of predicted values in the test set, and the percentage improvement in the MSE above the climatology and persistence baseline models. All three models were capable of producing an improved MSE above climatology and persistence baselines (see Table \ref{table2:cnn-results}, with the best results indicated in bold font). We then used the best performing model (DustNet) to visually evaluate its output against (unimputed) MODIS values. We inspected DustNet’s daily predictions for its ability to represent AOD spatially by mapping 28 consecutive days of predictions next to the corresponding data from MODIS (see supplementary Fig. \ref{figS6}). We looked for the model’s ability to capture the main dust generation sources, consistent AOD transport with prevailing winds, and correct distinctions of AOD accumulation between the ocean and land border. 

\begin{table}[!ht]
\caption{Normalised test results for three unique model architectures. Persistence and climatology baseline MSE’s of prediction to test data are presented below the table. The rows display results for total training time, time per iteration step and MSE for each kernel size of each model. The last column shows the percentage difference when compared to the climatological baseline.}\label{table2:cnn-results}%
\begin{tabular}{@{}lccccc@{}}
\toprule
CNN model & Training & Time & MSE & Prediction & Baseline\footnotemark[1] \\
 & time & per step & & time & improvement ({\%})\\
\midrule
Conv2D  & 13min40s & 34s & 0.001895 & 4.1s & $42.63\%$\\
U-NET    & 51min15s & 194s & 0.001904 & 18.6s & $42.36\%$\\
DustNet  & \textbf{7min41s} & \textbf{17s} & \textbf{0.00153} & \textbf{2.1s} & \textbf{53.68$\%$}\\
\botrule
\end{tabular}
\footnotetext{Baseline MSE:}
\footnotetext[1]{Climatology: 0.003303}
\footnotetext[2]{Persistence: 0.002992}
\end{table}

To analyse the errors of the best performing model, we rearranged Equation \ref{eq:normalisation} reverses normalisation of AOD predictions from each model:
\begin{equation}
    y_{ijt, denorm} = y_{ijt,pred}\ (y_{max} - y_{min}) + y_{min}
    \label{eq:de-normalisation}
\end{equation}
where y\textsubscript{pred} are the values predicted by the model, y\textsubscript{max} is the maximum and y\textsubscript{min} is the minimum AOD value from the training set. In this same manner, we used Equation \ref{eq:de-normalisation} to reverse normalisation of the climatology and persistence predictions. We then assessed each CNN model by calculating the MSE between values predicted by the model using the de-normalised AOD denoted as $\hat{A}$, and the corresponding values from the test set (“true”) AOD value devoted as \textit{A}. Here, we calculated a mean value along an axis of latitude N$_{lat}$ and longitude N$_{lon}$, of our spatial coordinates at each prediction time step t, where N$_{lat}$=31, N$_{lon}$=51 and N$_{t}$=1095, using Equation \ref{eq:MSE}:
\begin{equation}
    MSE = \frac{1}{N_{lat}N_{lon}N_t}\sum^{N_{lat}}_{i=1}\sum^{N_{lon}}_{j=1}\sum^{N_{t}}_{t=1} (\hat{A}_{ijt} - A_{ijt})^2
    \label{eq:MSE}
\end{equation}
We used this same process as described above to obtain the MSE for the climatology and persistence models. To ensure that model evaluation is only based on actually observed AOD values, all imputed AOD values were excluded from calculation of the MSE.

\paragraph*{DustNet evaluation metrics}
To compare predictions between the DustNet model, ground truth data from MODIS observations and the physics-based model (CAMS) fairly, we calculated the following metrics: mean bias error (MBE), RMSE, difference between RMSE’s ($\Delta$RMSE) and ACC. The metrics, defined below, follow a combination of notations from \cite{Bi2023} and \cite{Lam2023} adapted to spatial representation of temporally averaged values for each prediction day t (N$_{t}$=1095). All prediction values were first de-normalised using Equation \ref{eq:de-normalisation}. Subsequently, we compared the model predictions ($\hat{A}$) with raw (unimputed) MODIS data (mean of Aqua and Terra) denoted as \textit{A}. The climatological mean, denoted as \textit{A'}, corresponds to the long-term average of AOD values from MODIS (2003-2022).

\paragraph*{Spatial analysis} 
To analyse spatial characteristics of model performance, we calculated the temporal mean of model predictions (N$_{t}$ = 1095) at each location (lat,lon). This allowed us to calculate mean bias error (MBE) between the predicted AOD ($\hat{A}$) and MODIS ground truth (\textit{A}) for both DustNet and CAMS using Equation \ref{eq:MBE}.
\begin{equation}
    MBE_{spatial,ij} = \frac{1}{N_t}\sum_{t=1}^{N_t}(\hat{A}_{ijt} - A_{ijt})
    \label{eq:MBE}
\end{equation}
We also calculated the spatial root mean square error (RMSE$_{spatial}$) for each model using Equation \ref{eq:RMSE}. 
\begin{equation}
    RMSE_{spatial,ij} = \sqrt{\frac{1}{N_t}\sum_{t=1}^{N_t}(\hat{A}_{ijt} - A_{ijt})^2}
    \label{eq:RMSE}
\end{equation}
Calculating differences between RMSEs ($\Delta$RMSE) using Equation \ref{eq:delta-rmse} allowed us to reveal specific locations at which predictions from one model outperformed the other. 
\begin{equation}
    \Delta RMSE_{spatial,ij} = RMSE^{(CAMS)}_{spatial,ij} - RMSE^{(DustNet)}_{spatial,ij}
    \label{eq:delta-rmse}
\end{equation}
Additionally, we calculated the spatial distribution of Anomaly Correlation Coefficient (ACC, Equation \ref{eq:ACC}). Let $\hat{A}'$ be the anomaly of predicted AOD values ($\hat{A}$), and $A'$ the anomaly of observed (ground truth $A$) AOD values, where the anomalies are the differences from MODIS climatology values, then:
\begin{equation}
    ACC_{spatial,ij} = \frac{\sum_{t=1}^{N_t} \left[(\hat{A'}_{ijt} - \bar{A'}_{ijt}) \times (A'_{ijt} - \bar{A'}_{ijt})\right]} 
    {\sqrt{{\left[\sum_{t=1}^{N_t} (\hat{A'}_{ijt} - \bar{A'}_{ijt})^2\right] \times\left[ \sum_{t=1}^{N_t}(A'_{ijt} - \bar{A'}_{ijt})^2\right]}}}
    \label{eq:ACC}
\end{equation}
The ACC is a common measure of skill which assesses the quality of prediction, and highlights anomalies between forecast and observed values. By subtracting the climatological mean from both, prediction and verification, the ACC measures the quality of prediction without giving misleadingly high results caused by seasonal variations.
Refer to Fig. \ref{fig2}A-E and supplementary Fig. \ref{figS3} for graphed results of these calculations.

\paragraph*{Temporal analysis} 
To analyse the model’s predictions across different times, we calculated mean spatial AOD values for each prediction day. We also computed Pearson’s correlation coefficients (r), associated p-values, and coefficient of determination ($r^2$) using the SciPy statistical package v.1.12 for each prediction day (N=1095) of spatially averaged data (N$_{lat}$,N$_{lon}$=31,51). Corresponding results were calculated for both DustNet and CAMS forecasts with MODIS data and are plotted in supplementary Fig. \ref{figS1}. We have also adapted Equations \ref{eq:MBE} and \ref{eq:RMSE} to temporal representation by using Equation \ref{eq:MBE-temp} and \ref{eq:RMSE-temp}. 
\begin{equation}
    MBE_{temporal,t} = \frac{1}{N_{lat}N_{lon}}\sum^{N_{lat}}_{i=1}\sum^{N_{lon}}_{j=1}(\hat{A}_{ijt} - A_{ijt})
    \label{eq:MBE-temp}
\end{equation}
\begin{equation}
    RMSE_{temporal,t} = \sqrt{\frac{1}{N_{lat}N_{lon}}\sum^{N_{lat}}_{i=1}\sum^{N_{lon}}_{j=1}(\hat{A}_{ijt} - A_{ijt})^2}
    \label{eq:RMSE-temp}
\end{equation}
The graphed results of temporal calculations can be seen in Fig. \ref{fig4} and supplementary Fig. \ref{figS9:rmse-bias}.

\paragraph*{Justification of the selected points} 
In addition to spatial and temporal analyses, we focussed on four point locations to assess the model’s performance at the local scale. The locations, shown in supplementary Figure \ref{figS7:point-locations}, were selected on the basis of a different aerosol type contributing to the total AOD, as well as prevailing meteorological conditions. We chose the region around the Bod\'el\'e Depression in Chad (16.5\degree N, 16.5\degree E) for its dust generation capability and consistency of high mineral dust loading \cite{Washington2003}. Nouadhibou in Mauritania (20.5\degree N, 17\degree W) is located at the edge of western Africa, where hot, dry Saharan air meets cool and moist Atlantic air \cite{Carlson1972}. The temperature inversion creates a barrier for low horizontal flow of atmospheric dust, and instead forces an uplift of over 1.5km \cite{Prospero1972}. From this point atmospheric dust moves westward towards Central and South America at higher altitudes between 1.5km - 5km \cite{Kaufman2005}. To capture the transport of dust and fire smoke with southwestward winds towards South America \cite{Kaufman2005} we chose a location over the Atlantic Ocean in the Gulf of Guinea (4\degree N, 4\degree W). For the fourth location, we chose the second largest city in Nigeria and the capital of Kano State (11.5\degree N, 8.5\degree E). Kano City is on a direct pathway of seasonal dust plumes known locally as the Harmattan season. During boreal winter the wind direction shifts to southwestward direction and transports the sand storms generated from the Bod\'el\'e Depression towards Kano, where they are associated with a large increase in air pollution \cite{Anuforom2007, Schwanghart2008, Sunnu2008}.

\section*{Acknowledgements} We acknowledge NASA for producing, maintaining and releasing the MODIS AOD data which was used for training and comparison in this study. For these same reasons we acknowledge Copernicus Atmospheric Monitoring Service and ECMWF for their open release of CAMS AOD data.

For the purpose of open access, the author has applied a ‘Creative Commons Attribution (CC BY) licence to any Author Accepted Manuscript version arising from this submission.
\addcontentsline{toc}{section}{Acknowledgements}
\backmatter

\bmhead*{Funding} This work was supported by the UKRI Centre for Doctoral Training in Environmental Intelligence, Engineering and Physical Sciences Research Council Grant Reference: EP/S022074/1.

\bmhead*{Author contributions}Conceptualization: T.E.N., S.S., B.I.S., A.T.A.
	Methodology: T.E.N., S.S.
	Investigation: T.E.N.
	Visualization: T.E.N.
	Supervision: S.S., B.I.S., A.T.A.
	Writing—original draft: T.E.N.
	Writing—review and editing: T.E.N., S.S., B.I.S., A.T.A.

\bmhead*{Competing interests} The authors declare no competing interests.

\bmhead*{Code and data availability}\label{links}

The full code for each model (DustNet, U-NET and Conv2D) with structured input data were deposited \cite{Nowak2024compnote} and are available from Zenodo at \href{zenodo-models-analysis}{https://zenodo.org/records/10722953}. The repository includes all results from the DustNet model (output data), and Python code to replicate all statistical analysis to reproduce each figure included in this article. Pre-processed ERA5 and AOD data are deposited as NumPy files in Zenodo together with Python imputation code at \href{zenodo-pre-processed}{https://zenodo.org/records/10593152} \cite{Nowak2024dataset}.

Reanalysis of atmospheric features were downloaded from the Copernicus Climate Data Store collection `ERA5 hourly data on pressure levels from 1940 to present'. Unprocessed datasets are available from Copernicus Climate Change Services (C3S) Climate Data Store (CDS) at \href{era5}{https://cds.climate.copernicus.eu/cdsapp/}. Pre-processed ERA5 data is also included in the aforementioned Zenodo repository.

The AOD at 550nm Level 3 daily data for combined Dark Target and Deep Blue algorithms were retrieved from Moderate Resolution Imaging Spectroradiometer (MODIS) on both Aqua and Terra spacecraft. Both datasets are available from NASA’s Atmosphere Archive {\&} Distribution System (LAADS) Distributed Active Archive Center (DAAC). Both MOD08\_{D3} and MYD08\_{D3} files can be retrieved from \href{modis}{https://ladsweb.modaps.eosdis.nasa.gov/search/}.  Pre-processed AOD data is also included in the aforementioned Zenodo repository.

The forecast of AOD was downloaded from the Atmosphere Data Store of Copernicus Atmosphere Monitoring Service (CAMS). The total aerosol optical depth at 550nm from the Global atmospheric composition forecast for midday run with a 24hr lead-time can be obtained from \href{cams}{https://ads.atmosphere.copernicus.eu/\#!/home}.

\addcontentsline{toc}{section}{References}
\settocbibname{References}
\bibliography{sn-article}

\section*{Supplementary Figures}
\addcontentsline{toc}{section}{Supplementary Figures}

\begin{figure}[!ht]%
\centering
\includegraphics[width=1\textwidth, trim={0.5cm 3.7cm 0.5cm 2.7cm},clip]{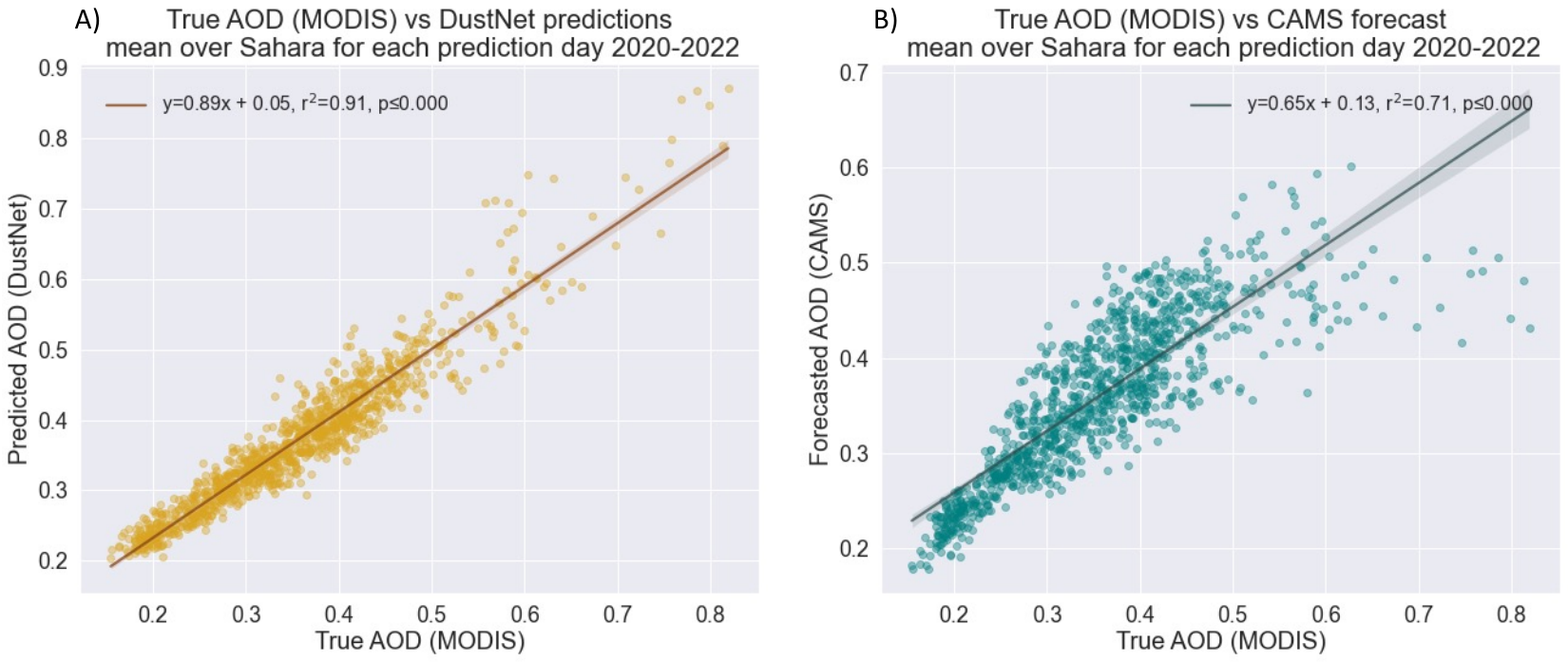}
\caption{Fig. S1: Daily spatial mean AOD (2020-2022) regressed between model predictions and MODIS data.S1. Linear regression with corresponding y equation, Pearson's r$^2$ and p values were calculated for daily spatial mean AOD over the Sahara for 2020 - 2022. Shown in \textbf{A)} AOD prediction results from DustNet correspond to MODIS data well with high r$^2$ = 0.91, and only a slight tendency to overestimate higher AOD. In \textbf{B)} the mean AOD forecasts from CAMS are shown to correspond with MODIS data well, r$^2$ = 0.71 though, with more frequent tendency to underestimate both low and high AOD values. Results from both predictions are highly significant with p$<0.0001$.}\label{figS1}
\end{figure}

\begin{figure}[!ht]%
\centering
\includegraphics[width=1\textwidth, trim={1cm 2.5cm 1cm 2.7cm},clip]{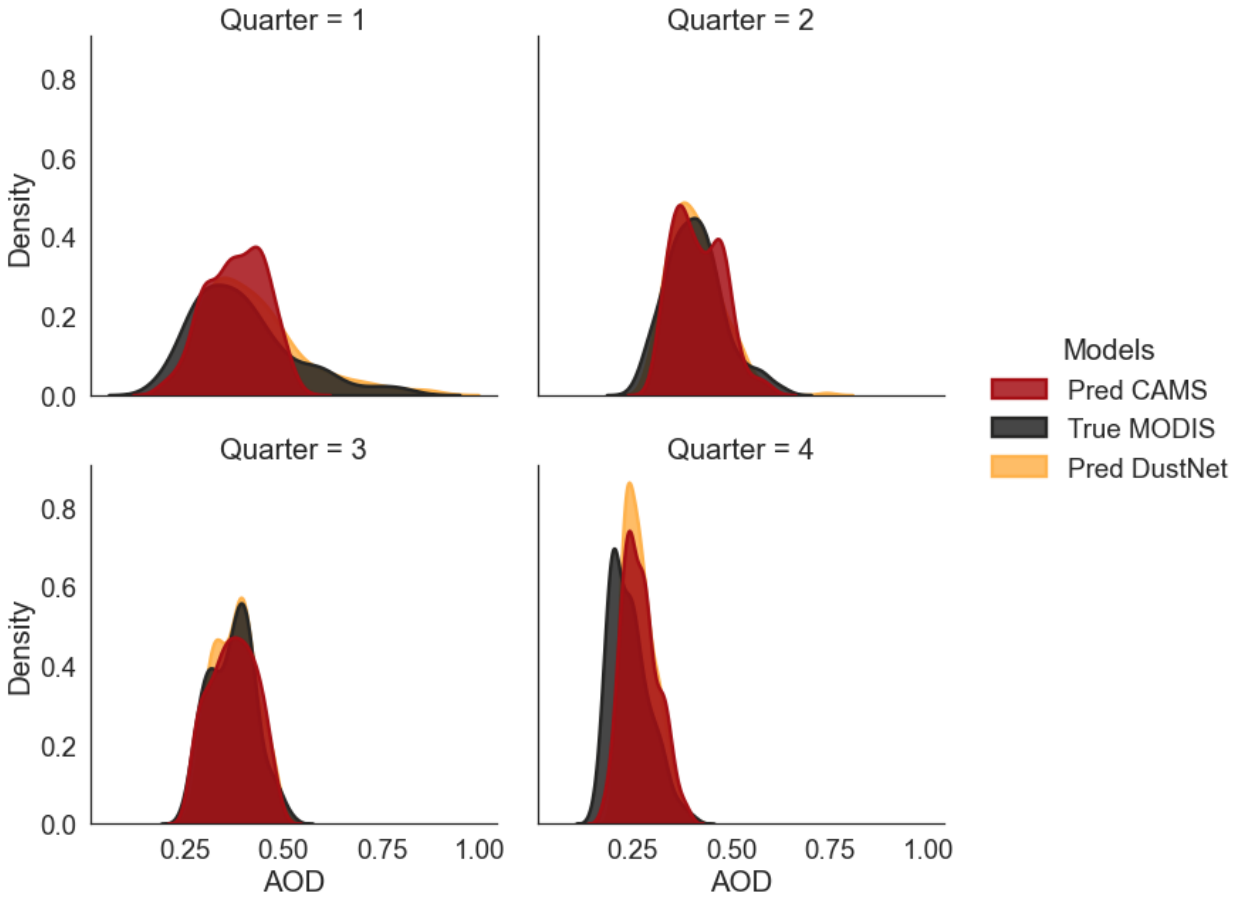}
\caption{Fig. S2: Seasonal mean distribution of daily AOD values. The data was averaged over the study region for the testing period of 2020-2022, and shows CAMS forecasts (red) DustNet predictions (yellow) and ground-true MODIS (black). The long tail, indicative of higher AOD values, is clearly missing in the CAMS distribution (red) for Quarter 1: January - March, while the lower AOD values are overestimated. The opposite is true for Quarter 4: October - December, where lower AOD values tend to be underestimated by both CAMS and DustNet in comparison to MODIS. Both models forecast fairly well during Quarter 2 and 3, although DustNet captures the bimodal distribution of AOD in Quarter 3 more skillfully than CAMS.}\label{figS2}
\end{figure}

\begin{figure}[!ht]%
\centering
\includegraphics[width=1\textwidth] {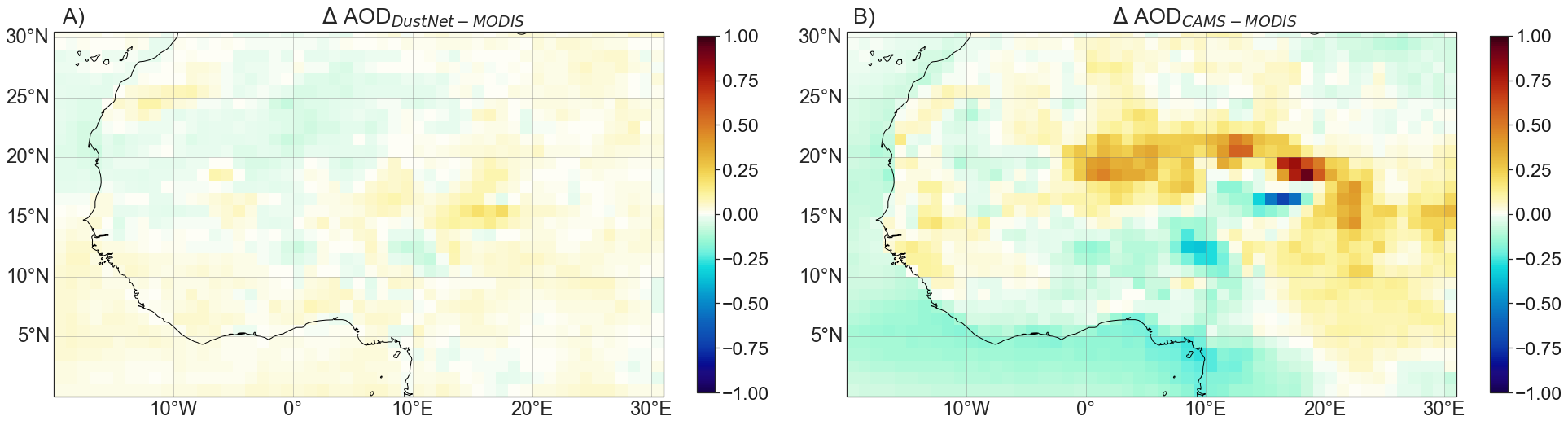}
\caption{Fig. S3: Bias of predictions \textbf{A)} DustNet and \textbf{B)} CAMS with respect to MODIS data. Note that the maximum bias produced by DustNet is 0.21 while the maximum bias for CAMS is 0.93.}\label{figS3}
\end{figure}

\begin{figure}[!ht]%
\centering
\includegraphics[width=0.87\textwidth, trim={1cm 1.5cm 1cm 1.5cm},clip]{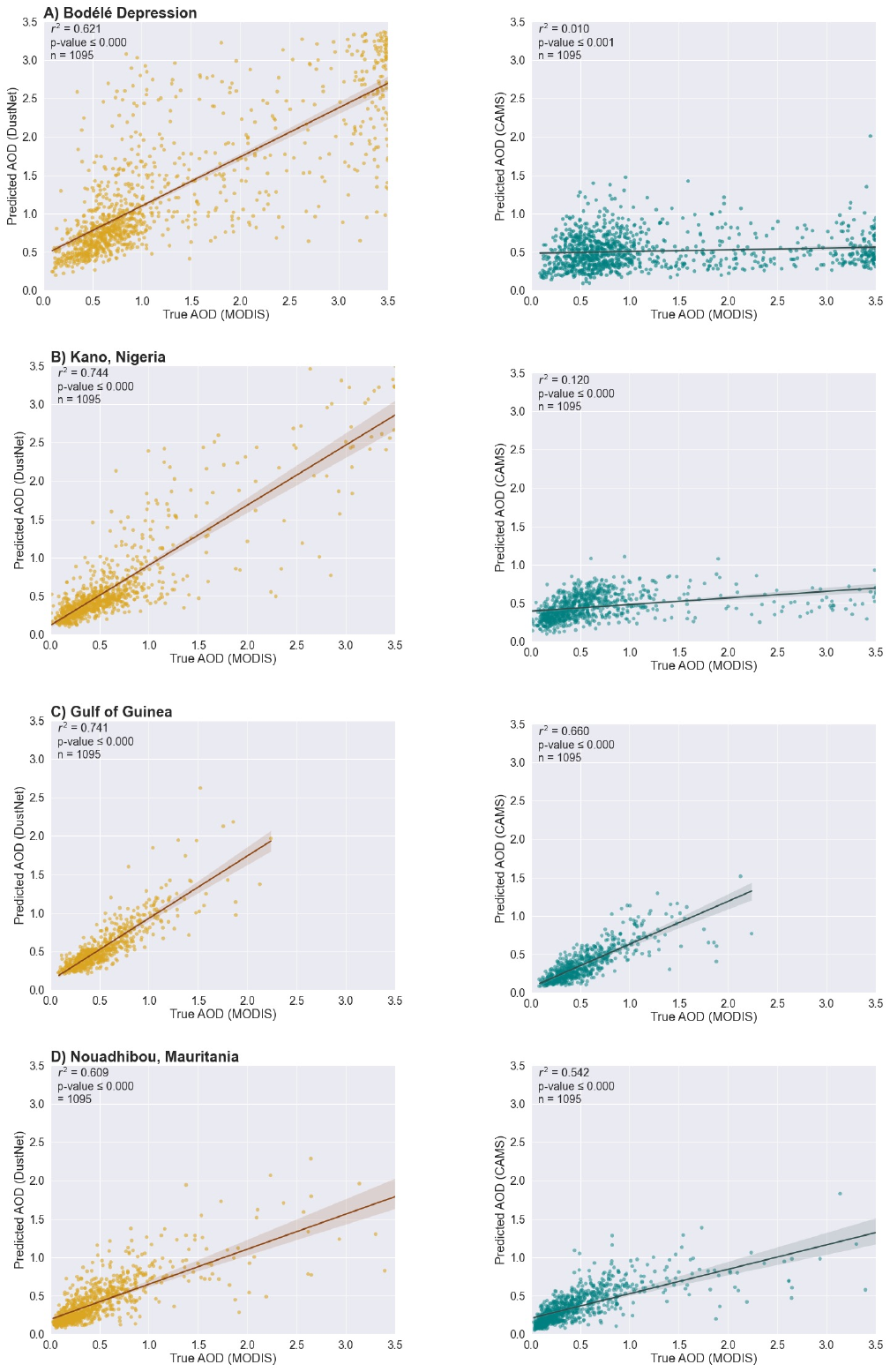}
\caption{Fig. S4: Scatter plot relationship between predicted mean AOD values (2020-2022) and MODIS data. Results for DustNet (\textbf{left panel}) and forecasts from CAMS (\textbf{right panel}) at four chosen locations show better agreement of DustNet predictions with MODIS data at each location. In \textbf{A)} the Bod\'el\'e Depression, Chad - highest source of dust in the Sahara - DustNet is significantly better than CAMS; \textbf{B)} Kano, Nigeria - the second most populous province \textbf{C)} Gulf of Guinea - over the ocean; and \textbf{D)} Nouadhibou, Mauritania - coastal location. }\label{figS4}
\end{figure}

\begin{figure}[!ht]%
\centering
\includegraphics[width=0.95\textwidth, trim={1cm 7.5cm 1cm 1.5cm},clip]{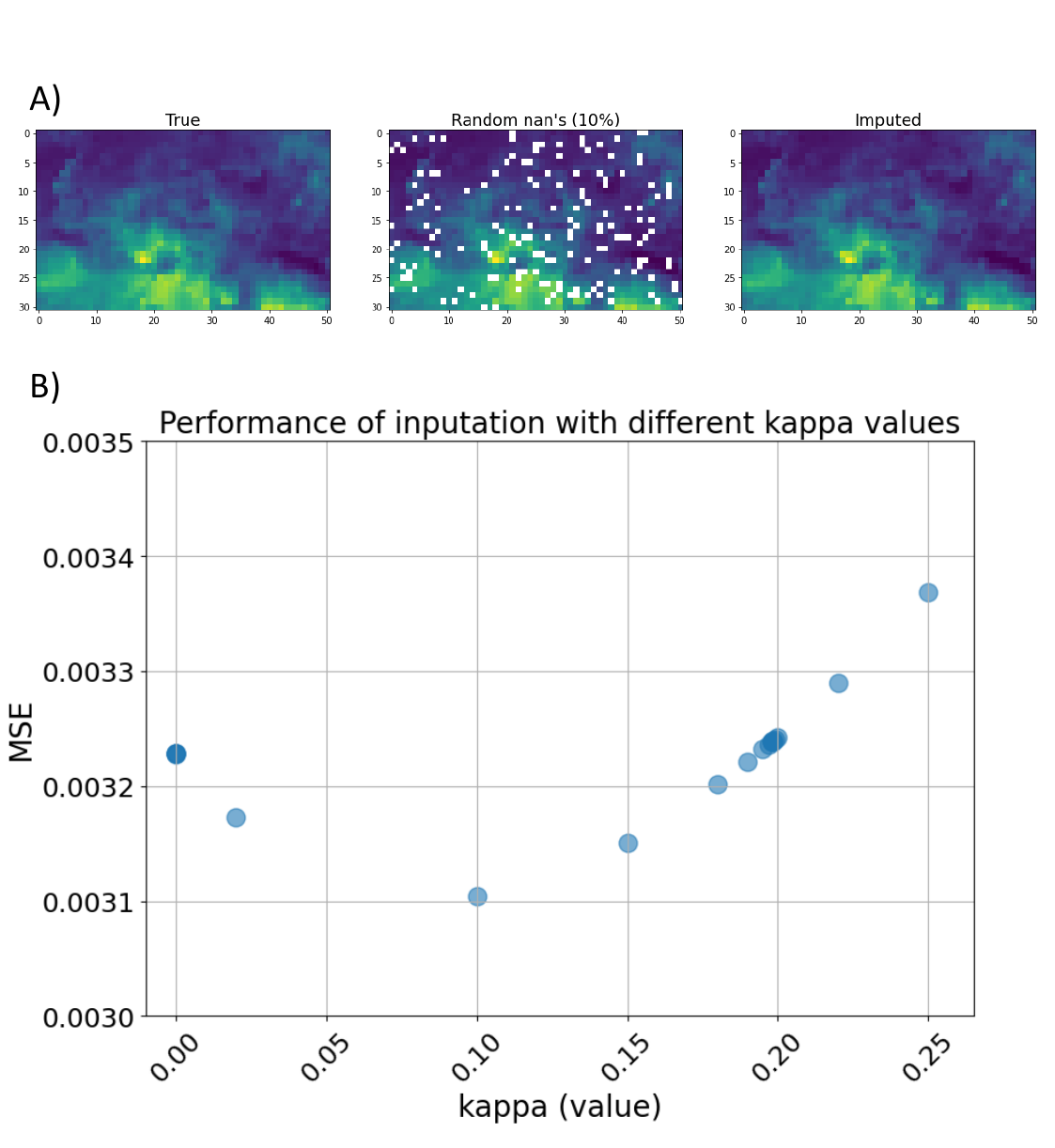}
\caption{Fig. S5: Performance validation of imputation code. Three visual inspections of imputation in \textbf{A)} show the "true" values on the left, randomly assigned $10\%$ of missing values in the middle, and imputed with Lattice Kriging method values in the right panel. In \textbf{B)} the results of MSEs between "true" and imputed values are displayed for different kappa values (Kriging hyperparameter) used during imputation. We tested 24 different kappa values ranging between 0.000001 and 0.1998) The MSE appears to be insensitive to changes in kappa values producing only marginal improvements in the overall MSE.}\label{figS5}
\end{figure}

\begin{figure}[!ht]%
\centering
\includegraphics[width=0.84\textwidth, trim={1.1cm 1.6cm 1.1cm 1.7cm},clip]{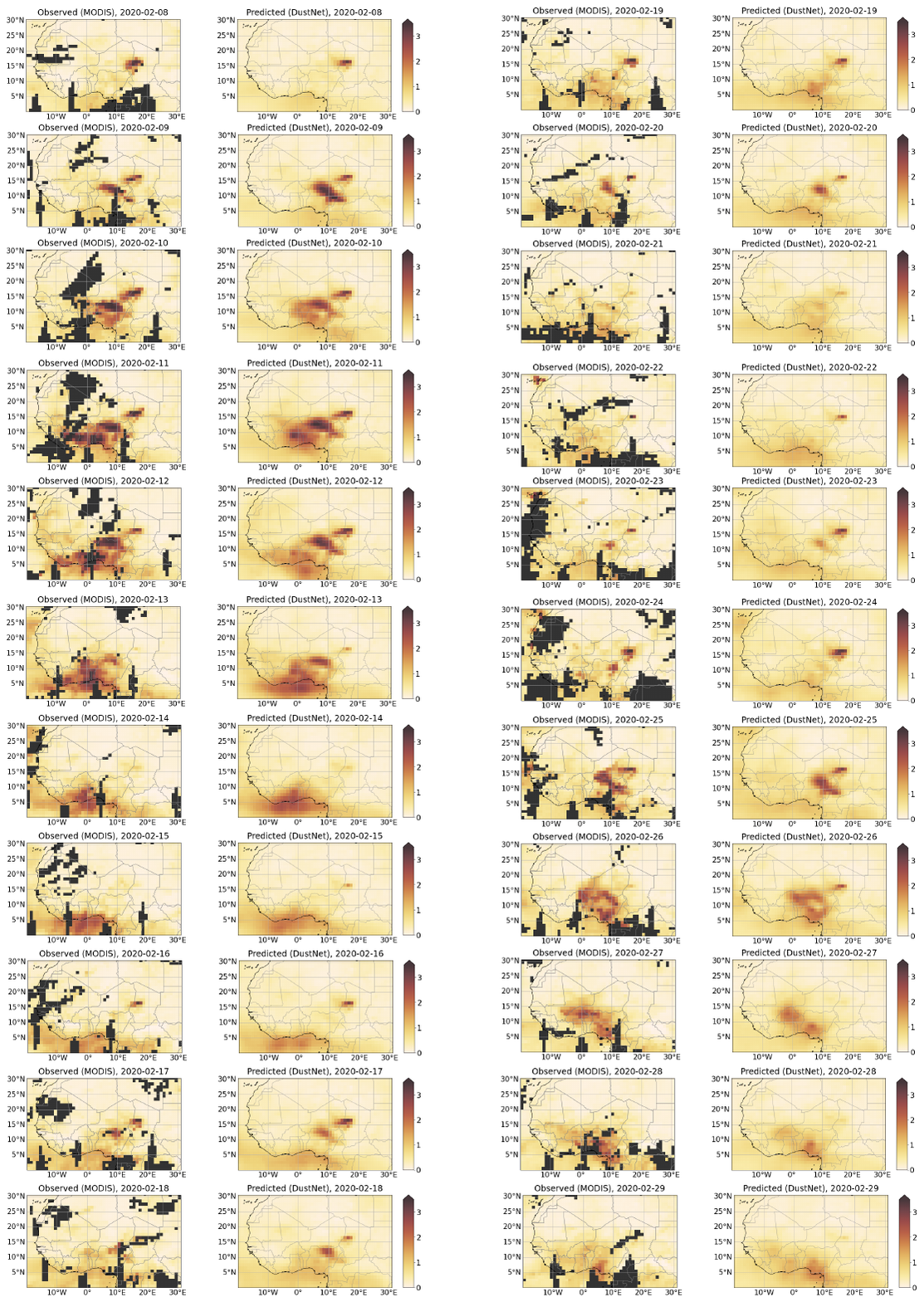}
\caption{Fig. S6: Comparison of daily AOD values as observed by MODIS (mean of Aqua and Terra) (\textbf{left panel} in both columns) and corresponding DustNet predictions (\textbf{right panel}) for selected continuous 3 weeks (22 days), from 8th - 29th February 2020. The dark grey colour in the MODIS maps represents missing values. Despite an initial assumption of heavy reliance on the past 5 days of AOD during training, DustNet presents a skillful ability to predict the next time-step (24-hr) which visibly differs from the last 5 days. This is evident on 13th -14th Feb and 21st Feb, where the AOD values start to decrease despite an increasing past trend. Similarly, prediction of an increasing AOD from 22nd - 26th Feb was captured, despite the previous 5 days of decreasing AOD. The south-western direction of aerosol transport during boreal winter is also skillfully captured (10th - 14th Feb), as is the position of the Bodélé Depression during dust generation episodes (22nd - 26th Feb), but without overly relying on this location as a constant dust source (27th - 29th Feb). }\label{figS6}
\end{figure}

\begin{figure}[htbp]
    \centering
    \includegraphics[width=1.0\textwidth]{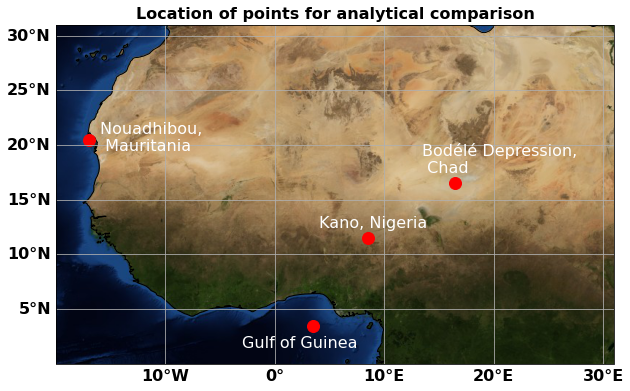}
    \caption{Fig. S7: Locations of selected grid points used to asses the model’s predictive accuracy on a local scale ($1\degree \times 1\degree\space$ grid size). The background image for the December view of Blue Marble is available from NASA https://visibleearth.nasa.gov/collection/1484/blue-marble?page=4 }
    \label{figS7:point-locations}
\end{figure}

\begin{figure}[htbp]
    \centering
    \includegraphics[width=1.0\textwidth]{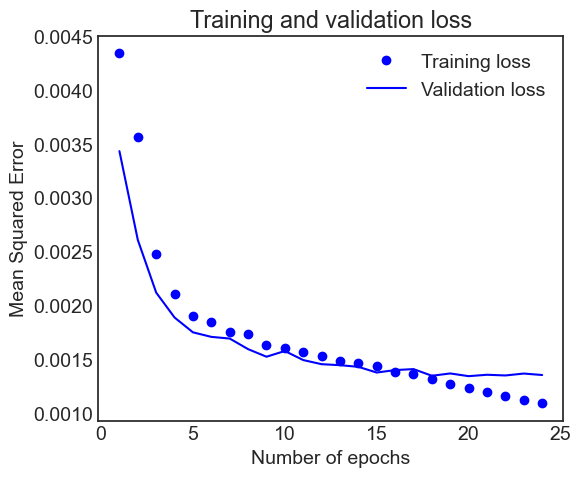}
    \caption{Fig. S8: Training and validation loss for the optimal model - DustNet. The model's architecture ensures Early Stopping is performed following the 4th iteration without any improvement in validation loss. Here, stopping occurred after 24 epochs and the model with the lowest ratio of training to validation loss was saved and used for predictions.}
    \label{figS8:train-valid}
\end{figure}

\begin{figure}[htbp]
    \centering
    \includegraphics[width=1.0\textwidth, trim={1cm 3.5cm 2.5cm 2.5cm},clip]{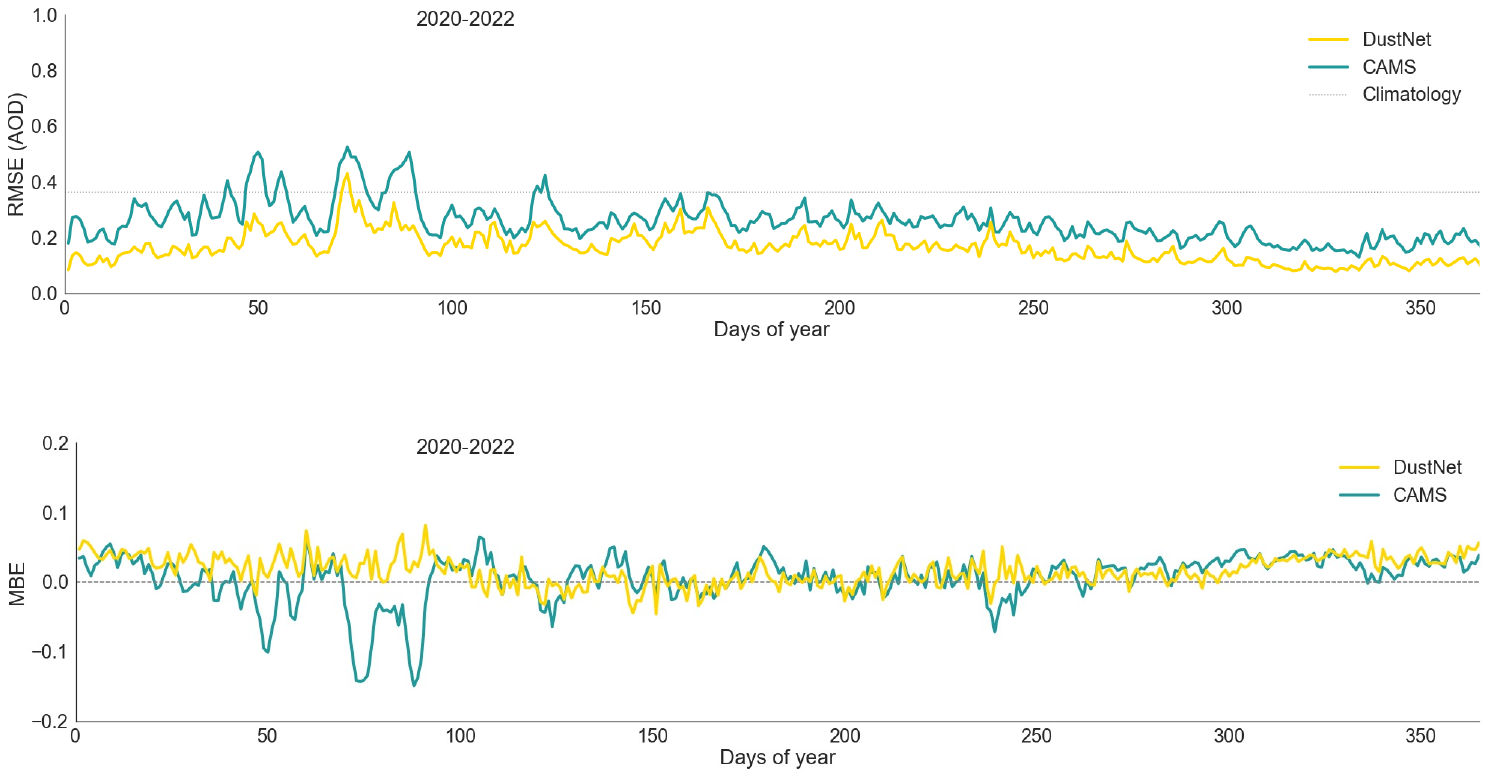}
    \caption{Fig. S9: In the \textbf{top panel} the temporal mean RMSE values for AOD predicted by DustNet (yellow) and CAMS (cyan) are shown. The dashed line represents the climatological mean from MODIS. At all time-steps the DustNet model predictions show smaller (better) errors than those produced by CAMS. The \textbf{lower panel} shows the temporal mean bias errors (MBE) from the DustNet predictions (yellow) and CAMS (cyan). Here, the DustNet bias fluctuates close to zero more often than the bias produced by CAMS.}
    \label{figS9:rmse-bias}
\end{figure}






\end{document}